\documentclass[aps,physrev,reprint,superscriptaddress]{revtex4-2}

\usepackage{amsmath,amssymb}
\usepackage{graphicx}
\usepackage{url}

\begin{document}

\title{Finite-Response Complementarity in Fluctuation Constraints on Climate Sensitivity}

\author{Zaibo Zhao}
\affiliation{Yunnan Key Laboratory of Complex Systems and Brain-Inspired Intelligence, Kunming University of Science and Technology, Kunming, Yunnan 650500, China}
\affiliation{Faculty of Science, Kunming University of Science and Technology, Kunming 650500, China}
\author{Yongwen Zhang}
\email{zhangyw@kust.edu.cn}
\affiliation{Yunnan Key Laboratory of Complex Systems and Brain-Inspired Intelligence, Kunming University of Science and Technology, Kunming, Yunnan 650500, China}
\affiliation{Faculty of Science, Kunming University of Science and Technology, Kunming 650500, China}

\date{\today}

\begin{abstract}
Equilibrium climate sensitivity (ECS) is a zero-frequency susceptibility, whereas historical global-mean temperature variability samples a finite, forced projection of the climate system. We test whether information missed by a scalar fluctuation-memory coordinate reappears in a finite CO$_2$ response. After conditioning both ECS and the finite response on \(\Psi\), CMIP5 residuals are nearly uncoupled (\(C_5=0.154\)), whereas CMIP6 shows strong complementarity (\(C_6=0.593\), \(p=0.00288\)). A two-mode stochastic response model attributes this contrast to hidden-response spread that is visible in the finite response but poorly projected onto \(\Psi\). The CMIP6 residual direction defines a first-order correction and yields a HadCRUT5 conditional ECS estimate centered at \(3.00\,\mathrm K\). Thus the weakened CMIP6 fluctuation constraint does not imply that susceptibility information is lost: part of it is recovered through a complementary finite-response projection.
\end{abstract}

\maketitle

\section{Introduction}

Equilibrium climate sensitivity (ECS), the long-time global-mean surface-temperature response to doubled atmospheric CO$_2$, is proportional to a zero-frequency susceptibility and remains a central uncertainty in climate projection \cite{bib1,bib2}. Emergent constraints seek observable predictors of intermodel ECS spread, but their credibility rests on physical mechanisms and testable assumptions rather than correlation alone \cite{bib3,bib4}. \citet{bib5} proposed a fluctuation-memory statistic \(\Psi\), computed from historical global-temperature variability, that appeared to constrain ECS in CMIP5. In response-theory language, \(\Psi\) is a scalar projection of a forced historical trajectory; it can constrain the zero-frequency susceptibility only if the response-relevant degrees of freedom lie in, are small relative to, or are aligned with that projection.

Linear response theory provides the physical reason why fluctuation statistics might constrain response. In the equilibrium fluctuation-dissipation theorem (FDT), the susceptibility to a weak perturbation is determined by correlation functions of unperturbed equilibrium fluctuations \cite{bib6}. Climate applications imported this idea through stochastic climate modelling, fluctuation-dissipation estimates, and tests of FDT operators in general circulation models \cite{bib7,bib8,bib9,bib10}. The historical record used in emergent constraints, however, is not the equilibrium object of that theorem. It is finite, externally forced, and observed through a low-dimensional projection, usually global-mean surface temperature (GMST). Nonequilibrium climate-response theory treats variability, change, sensitivity, response, and predictability as linked but distinct aspects of a forced dissipative system \cite{bib11,bib12}, while Ruelle response theory and projection formalisms show how susceptibilities, memory, and effective noise arise when unresolved degrees of freedom are projected out \cite{bib13,bib14,bib15,bib16}. The relevant closure problem is therefore which susceptibility components are captured by a scalar fluctuation coordinate, and where the remaining response information is projected.

This variability constraint became a useful test case because its physical interpretation was contested. Subsequent studies questioned its robustness, possible forced-signal content, and transfer beyond CMIP5 \cite{bib4,bib17,bib18}; direct CMIP6 tests showed a marked weakening \cite{bib19}, and \citet{bib20} traced the difference to a scalar assumption that worked approximately in CMIP5 but failed in CMIP6. This shift is physically plausible because CMIP6 exhibits larger spread in effective sensitivity and stronger feedback heterogeneity, particularly associated with cloud feedbacks and pattern effects \cite{bib21,bib22}. Thus the historical scalar fluctuation coordinate may retain susceptibility information without closing the full forced-response susceptibility.

Here we recast the problem as fluctuation-response complementarity under projection. The purpose of introducing a finite CO$_2$ response coordinate \(R\), estimated without ECS labels, is not to add an empirical predictor, but to test where response information not represented by \(\Psi\) is projected. We therefore ask whether the ECS residual left by \(\Psi\) is organized by the residual of \(R\). Such residual transmission would indicate that a finite forced trajectory contains response-relevant information outside the scalar fluctuation-memory projection.

\section{Data and Methods}

\subsection{Data and preprocessing}

The main analysis uses annual global-mean near-surface air temperature (tas) from available CMIP5 and CMIP6 historical simulations. A model enters the analysis if it has a historical annual global-mean tas series, an ECS label, and a finite-response coordinate estimated from the historical forced trajectory. The resulting sample contains 18 CMIP5 models and 23 CMIP6 models. HadCRUT5 is used for the observation-facing diagnostic and is processed through the same windowing and response-feature pipeline. CMIP5 and CMIP6 simulations were obtained from the Earth System Grid Federation archive \cite{bib23,bib24}; HadCRUT5 follows \citet{bib25}. Effective radiative forcing histories for CO$_2$, anthropogenic aerosol, and natural forcing are applied as common forcing components. The fitted response feature is therefore a common-forcing response coordinate inferred from each temperature trajectory, not a model-specific forcing reconstruction or an energy-budget inversion.

Throughout, \(T_i(t)\) denotes annual GMST for model \(i\), \(E_i\equiv\mathrm{ECS}_i\simeq F_{2\times\mathrm{CO}_2}\chi_i(\infty)\) is its long-time response, and \(R_i\equiv R_{\mathrm{CO}_2,i}\simeq F_{2\times\mathrm{CO}_2}\chi_i(\tau)\) is its finite-time response. Here \(F_{2\times\mathrm{CO}_2}\) is the effective radiative forcing for CO$_2$ doubling and \(\chi_i\) is the projected global-temperature susceptibility.

\subsection{Fluctuation-memory coordinate}

For model \(i\), the fluctuation-memory coordinate is computed in rolling 55-year historical windows:
\begin{equation}
\Psi_i
=
\left\langle
\frac{\sigma_{i,w}}{\sqrt{-\log \alpha_{1,i,w}}}
\right\rangle_w ,
\label{eq:psi}
\end{equation}
where \(\sigma_{i,w}\) and \(\alpha_{1,i,w}\) are respectively the detrended standard deviation and one-year lag autocorrelation in window \(w\), and \(\langle\cdot\rangle_w\) denotes the mean across valid windows. This keeps the calculation close to the historical finite-window construction of \citet{bib5}. The analysis does not assume that historical \(\Psi\) is a pure internal-variability statistic; linear detrending reduces the leading warming trend but does not perfectly separate internal variability from forced structure.

\subsection{Finite CO$_2$ response coordinate}

The finite-response coordinate \(R_{\mathrm{CO}_2}\) is estimated without ECS labels. For each model, a scalar forced ARX/state-space model is fitted to annual temperature using common effective radiative forcing components:
\begin{subequations}
\label{eq:arx-response}
\begin{align}
\begin{split}
T_{t+1}
&=aT_t+\beta_{\mathrm{CO}_2}F_{\mathrm{CO}_2,t+1}
+\beta_{\mathrm{aer}}F_{\mathrm{aer},t+1}\\
&\quad+\beta_{\mathrm{nat}}F_{\mathrm{nat},t+1}
+c+\epsilon_{t+1},
\end{split}\\
R_{\mathrm{CO}_2}
&=
T_{\mathrm{step}}(55)-T_{\mathrm{control}}(55).
\end{align}
\end{subequations}
After fitting, a CO$_2$-doubling-equivalent step forcing is applied to the fitted model and the response is evaluated at the same default 55-year horizon used for the fluctuation windows. A conservative forced Neural ODE estimator is used as a nonlinear response-control analysis in Supplementary Note 3 and Supplementary Fig. S2; it gives nearly collinear response coordinates with the ARX estimator.

\subsection{Residual complementarity framework and significance}

If the scalar fluctuation projection were sufficient, the projected long-time susceptibility would be described by \(E\approx E_0(\Psi)\). More generally, the long-time susceptibility can also depend on a hidden forced-response coordinate \(\eta\) that is not fully resolved by \(\Psi\), so \(E=E(\Psi,\eta)\). The same hidden coordinate can leave a measurable trace in the finite historical response, \(R=R(\Psi,\eta)\). Thus \(R\) provides an observable projection through which forced-response information can complement the scalar fluctuation-memory coordinate.

Within each ensemble \(g\), here CMIP5 or CMIP6, the scalar references \(E_{0,g}(\Psi)=\mathbb E_g[E|\Psi]\) and \(R_{0,g}(\Psi)=\mathbb E_g[R|\Psi]\) are estimated independently by ordinary least squares. The lowest-order linear form is used in the main analysis, with nonlinear alternatives tested in Supplementary Note 5. The residuals are
\begin{subequations}
\label{eq:residual-definition}
\begin{align}
\Delta E_{i,g}
&=E_i-E_{0,g}(\Psi_i),\\
\Delta R_{i,g}
&=R_i-R_{0,g}(\Psi_i).
\end{align}
\end{subequations}
Their alignment is measured by the finite-response complementarity coefficient
\begin{equation}
C_g=\operatorname{corr}_g(\Delta E,\Delta R).
\label{eq:finite-response-complementarity}
\end{equation}
The marginal \(\Psi\)-ECS relation measures the scalar constraint itself; \(C_g\) instead tests whether the part of ECS left unexplained by \(E_{0,g}(\Psi)\) is aligned with the finite-response residual. Thus \(C_g\simeq0\) means that \(R\) adds little after conditioning on \(\Psi\), whereas positive \(C_g\) means that complementary ECS information reappears along \(\Delta R\).

To use this alignment as a correction, the residuals must be dominated locally by a common unresolved forced-response coordinate \(\delta\eta\). A first-order expansion at fixed \(\Psi\) gives
\begin{subequations}
\label{eq:gamma-operator}
\begin{align}
\Delta E&\approx (\partial_\eta E)_\Psi\,\delta\eta,&
\Delta R&\approx (\partial_\eta R)_\Psi\,\delta\eta,\label{eq:hidden-linearization}\\
\Delta E&\approx \Gamma_\eta\,\Delta R,&
\Gamma_\eta&=\frac{(\partial_\eta E)_\Psi}{(\partial_\eta R)_\Psi}.
\label{eq:gamma-elimination}
\end{align}
\end{subequations}
Real ensembles can contain several unresolved coordinates (Supplementary Note 6) and model-dependent finite-time kernels. We therefore estimate the ensemble-level effective residual operator
\begin{equation}
\Gamma_g=
\frac{\operatorname{Cov}_g(\Delta E,\Delta R)}
{\operatorname{Var}_g(\Delta R)}
=C_g\frac{\sigma_{\Delta E,g}}{\sigma_{\Delta R,g}}.
\label{eq:residual-operator}
\end{equation}
Here \(C_g\) measures residual alignment after scalar conditioning, whereas \(\Gamma_g\) gives the conversion scale when such alignment is present. Permutation significance uses a residual-pairing null in which \(\Delta E\) is held fixed while \(\Delta R\) is randomly reassigned across model labels within the same ensemble. The two-sided permutation probability is computed from 50,000 random reassignments. Bootstrap, model-omission checks, and alternative \(R_0(\Psi)\) references are reported in Supplementary Notes 4 and 5 and Supplementary Figs. S3-S5.

\subsection{Two-mode stochastic response model}

To interpret residual complementarity, we use a controlled projection experiment rather than a comprehensive climate emulator. The model contains an observed GMST-like coordinate \(T\), a hidden slow response state \(z\), and common forcing \(F\):
\begin{equation}
\begin{aligned}
\dot T&=-(\lambda_i/C_{T,i})T+(\kappa_i/C_{T,i})z+(b_i/C_{T,i})F+\xi_T,\\
\dot z&=-(\mu_i/C_{z,i})z+(q_i/C_{z,i})F+\xi_z.
\end{aligned}
\label{eq:toy-dynamics}
\end{equation}
With \(B_i=b_i/\lambda_i\) and hidden susceptibility amplitude \(H_i=\kappa_iq_i/(\lambda_i\mu_i)\), its long-time and finite-time responses are
\begin{subequations}
\label{eq:toy-responses}
\begin{align}
E_i&=F_{2\times\mathrm{CO}_2}(B_i+H_i),\label{eq:toy-long-response}\\
R_i(\tau)&=F_{2\times\mathrm{CO}_2}
\left[B_iG_T(\tau)+H_iG_z(\tau)\right].\label{eq:toy-finite-response}
\end{align}
\end{subequations}
The theoretical hidden coordinate \(\eta\) is the cross-model strength of the forced pathway \(F\to z\to T\), rather than the instantaneous state \(z(t)\). Its component unresolved by the scalar fluctuation coordinate is represented by
\begin{equation}
\begin{aligned}
\log H_i&=\log H_0+s\left(\rho_\Psi u_i+\sqrt{1-\rho_\Psi^2}\,v_i\right),\\
s_\perp&=s\sqrt{1-\rho_\Psi^2},
\end{aligned}
\label{eq:hidden-spread}
\end{equation}
where \(u_i\) is the standardized scalar fluctuation coordinate, \(v_i\) is an independent disorder direction, and \(\rho_\Psi\simeq\operatorname{corr}(\log H,\Psi)\) is the projected observability of hidden susceptibility by \(\Psi\).

The finite-response residual contains both a response-amplitude component \(r_{\rm amp}\), which can change the long-time susceptibility, and a kinetic component \(r_{\rm kin}\), generated by finite-time recovery-factor variation. Writing \(\Delta R=r_{\rm amp}+r_{\rm kin}\), the response-information ratio \(\mathcal Q=\|r_{\rm amp}\|^2/\|r_{\rm kin}\|^2\) predicts
\begin{equation}
C_{\rm pred}\simeq\sqrt{\frac{\mathcal Q}{1+\mathcal Q}}.
\label{eq:c-prediction}
\end{equation}
When \(r_{\rm amp}\) dominates, the same amplitude residual affects both \(R\) and \(E\), so \(C_{\rm pred}\) is large; when \(r_{\rm kin}\) dominates, \(R\) varies mainly because of finite-time adjustment differences and carries little residual information about \(E\). Supplementary Note 7 gives the finite-time kernels, parameter distributions, representative ensembles, and the derivation of Eq.~(\ref{eq:c-prediction}).

\subsection{HadCRUT5 conditional diagnostic}

HadCRUT5 provides the observable coordinate pair \((\Psi_\oplus,R_\oplus)\), but not the unknown long-time response \(E_\oplus\). Using the CMIP6 scalar references and residual operator, we define the observation-facing corrected coordinate
\begin{equation}
\hat E_{\rm corr}
=E_0^{(6)}(\Psi)+\Gamma_6[R-R_0^{(6)}(\Psi)].
\label{eq:corrected-estimator}
\end{equation}
Uncertainty is propagated by Monte Carlo sampling of HadCRUT5 rolling-window \(\Psi_\oplus\), ARX coefficient-covariance samples for \(R_\oplus\), and CMIP6 bootstrap draws of \(E_0\), \(R_0\), and \(\Gamma\). The corrected-coordinate samples are passed through the CMIP6 conditional density
\begin{equation}
\label{eq:conditional-density}
\begin{aligned}
p(E_\oplus|\mathcal D_\oplus,\mathcal M_6)
&=
\int
p_6(E_\oplus|\hat E_{{\rm corr},\oplus})\\
&\quad\times
p(\hat E_{{\rm corr},\oplus}|\mathcal D_\oplus,\mathcal M_6)
\,d\hat E_{{\rm corr},\oplus},
\end{aligned}
\end{equation}
where \(\mathcal D_\oplus\) denotes HadCRUT5 and common forcing data, and \(\mathcal M_6\) denotes the CMIP6 calibration ensemble. The quoted interval is an equal-tailed 66\% conditional diagnostic interval, reported to align with the likely-range convention used in IPCC assessments. The full Monte Carlo implementation and sensitivity tests are given in Supplementary Note 8 and Supplementary Fig. S7.

\section{Results}

\subsection{Coordinate-level switch across CMIP generations}

We first compare the two projected coordinates with ECS across model generations. The fluctuation-memory coordinate \(\Psi\) combines detrended fluctuation amplitude and one-year memory in 55-year historical windows, whereas \(R_{\mathrm{CO}_2}\) is the 55-year finite CO$_2$ step response estimated from annual temperature and common forcing histories without using ECS labels (Data and Methods).

\begin{figure*}[t]
\centering
\includegraphics[width=\textwidth]{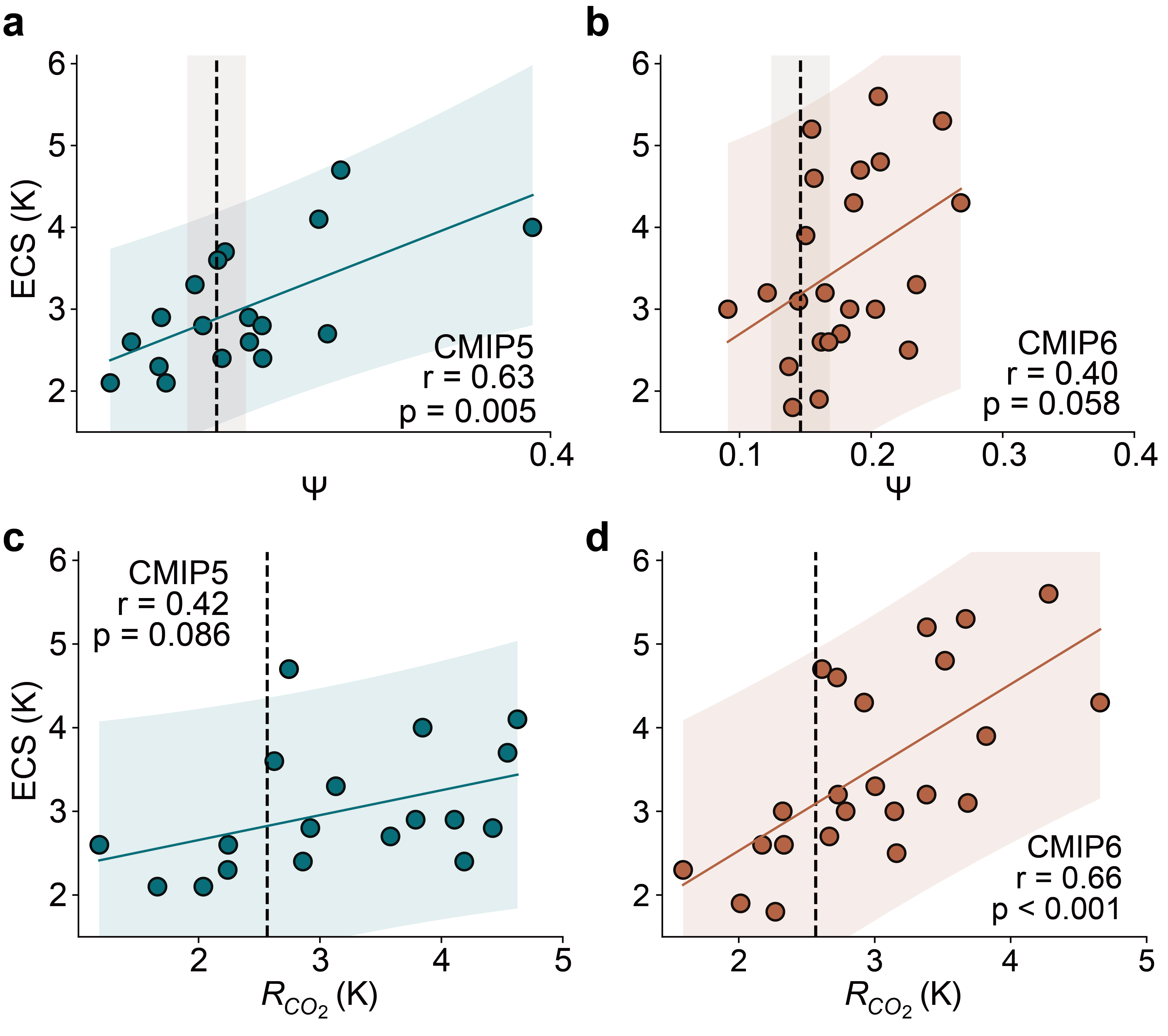}
\caption{\label{fig:coordinate-switch} \textbf{Coordinate-level empirical signature.} \textbf{a}, CMIP5 ECS versus the fluctuation-memory coordinate \(\Psi\). \textbf{b}, The same \(\Psi\)-ECS relationship in CMIP6. \textbf{c}, CMIP5 ECS versus the finite CO$_2$ response coordinate \(R_{\mathrm{CO}_2}\). \textbf{d}, The same finite-response relationship in CMIP6. Colored shaded regions are model-calibration predictive bands. Dashed vertical markers show the HadCRUT5 coordinate locations; gray vertical bands in \textbf{a},\textbf{b} show HadCRUT5 rolling-window spread in \(\Psi\).}
\end{figure*}

In the CMIP5 near-surface air-temperature archive, the fluctuation-memory coordinate \(\Psi\) is positively related to ECS (Fig. \ref{fig:coordinate-switch}a; \(r=0.63\), \(p=0.005\)). The same scalar relationship is weaker in CMIP6 (Fig. \ref{fig:coordinate-switch}b; \(r=0.40\), \(p=0.058\)), although the HadCRUT5 \(\Psi\) range remains within the CMIP6 ensemble range. The finite-response coordinate shows the complementary pattern: \(R_{\mathrm{CO}_2}\) is only weakly related to ECS in CMIP5 (Fig. \ref{fig:coordinate-switch}c; \(r=0.42\), \(p=0.086\)), but becomes a strong marginal coordinate in CMIP6 (Fig. \ref{fig:coordinate-switch}d; \(r=0.66\), \(p<10^{-3}\)). Thus Fig. \ref{fig:coordinate-switch} motivates the residual test of whether ECS information not organized by \(\Psi\) reappears in the finite-response projection. Additional coordinate diagnostics are given in Supplementary Notes 1-3 and Supplementary Figs. S1 and S2.

\subsection{Finite-response complementarity after scalar conditioning}

We next condition both the long-time and finite-time response coordinates on \(\Psi\) within each ensemble and evaluate the residual alignment \(C_g\) and conversion scale \(\Gamma_g\) defined in Eqs.~(\ref{eq:finite-response-complementarity}) and (\ref{eq:residual-operator}). This separates finite-response complementarity from the marginal \(\Psi\)-ECS relationship.

\begin{figure*}[t]
\centering
\includegraphics[width=\textwidth]{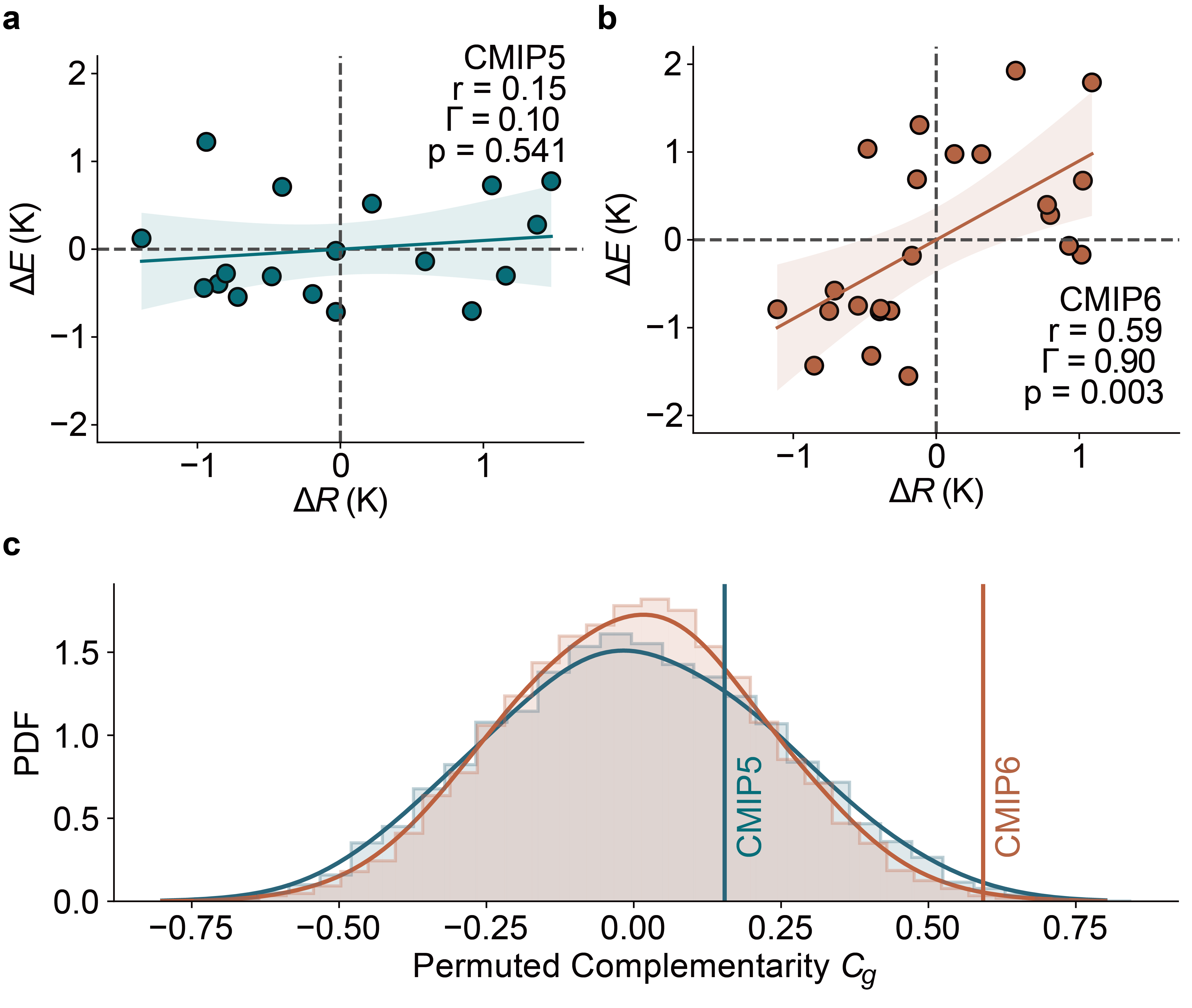}
\caption{\label{fig:self-closure} \textbf{Finite-response complementarity after scalar conditioning.} \textbf{a}, CMIP5 residual ECS, \(\Delta E_5\), versus residual finite response, \(\Delta R_5\), after subtracting \(E_{0,5}(\Psi)\) and \(R_{0,5}(\Psi)\). \textbf{b}, The same residual projection in CMIP6. Pale bands are 95\% confidence intervals for the fitted regression mean. \textbf{c}, Permutation null distributions for the finite-response complementarity coefficient \(C_g=\operatorname{corr}(\Delta E,\Delta R)\), with observed CMIP5 and CMIP6 values marked.}
\end{figure*}

CMIP5 shows little residual complementarity (Fig. \ref{fig:self-closure}a): \(\Delta E_5\) has no systematic dependence on \(\Delta R_5\), with \(C_5=0.154\), \(\Gamma_5=0.099\), and \(p=0.541\). CMIP6 behaves differently (Fig. \ref{fig:self-closure}b). Models with larger finite-response residuals also have larger ECS residuals, giving \(C_6=0.593\), \(\Gamma_6=0.899\), and \(p=0.00288\). The permutation nulls in Fig. \ref{fig:self-closure}c place the CMIP5 value inside the null distribution but the CMIP6 value in its upper tail. The central empirical result is therefore not simply that the marginal \(\Psi\)-ECS slope changes in CMIP6; rather, the information not carried by the fluctuation-memory coordinate is systematically projected into an observable finite-response residual. Additional robustness checks are shown in Supplementary Note 4 and Supplementary Figs. S3-S5.

\subsection{A minimal stochastic response mechanism}

The two-mode stochastic response model defined in Data and Methods provides a controlled test of whether unresolved response-amplitude spread can produce the residual complementarity seen in CMIP6, and whether kinetic/time-scale variation can obscure it.

\begin{figure*}[t]
\centering
\includegraphics[width=\textwidth]{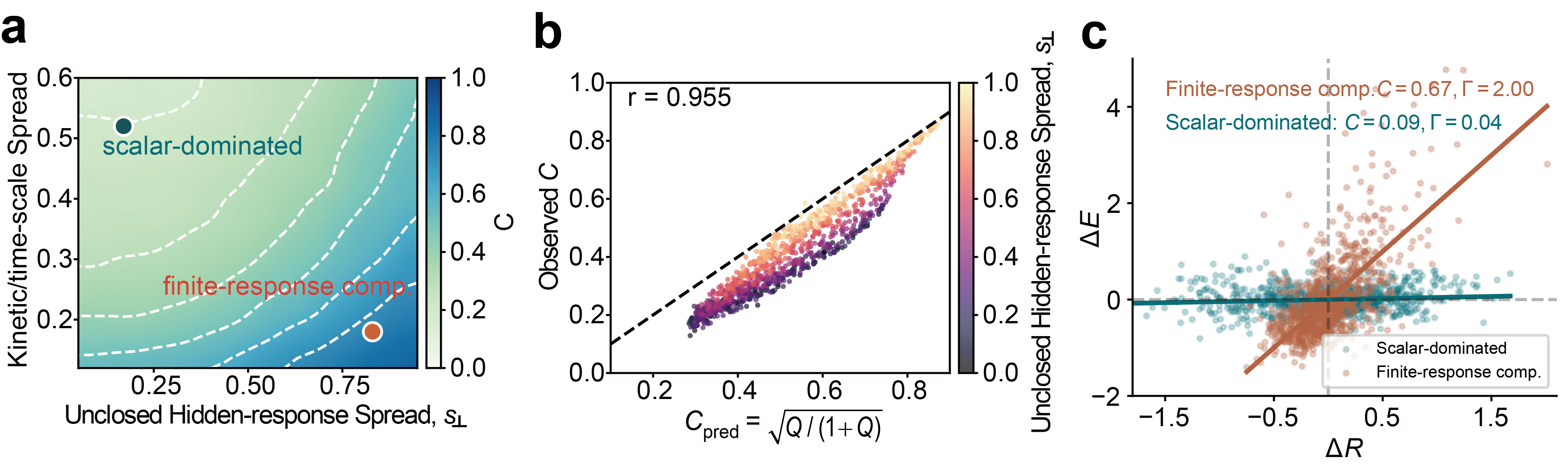}
\caption{\label{fig:toy-mechanism} \textbf{Minimal projected stochastic mechanism.} \textbf{a}, Seed-averaged finite-response complementarity coefficient \(C\) in the two-mode stochastic model as a function of unresolved hidden-response spread \(s_\perp\) and kinetic/time-scale spread. Markers show representative scalar-dominated and finite-response-complementary ensembles. \textbf{b}, The response-information prediction \(C_{\rm pred}=\sqrt{\mathcal Q/(1+\mathcal Q)}\) against the observed complementarity coefficient; points are colored by \(s_\perp\). \textbf{c}, Complementarity examples for the two marked ensembles.}
\end{figure*}

Fig. \ref{fig:toy-mechanism}a maps \(C\) across the two control directions in synthetic ensembles generated from the two-mode model. Increasing \(s_\perp\) supplies unresolved response-amplitude spread, while increasing kinetic/time-scale spread adds recovery-factor contamination to \(\Delta R\). Finite-response complementarity is therefore strongest in the large-\(s_\perp\), weak-contamination regime, and weakest when the hidden spread is small or \(\Delta R\) is dominated by recovery-factor variability. Fig. \ref{fig:toy-mechanism}b compares the observed \(C\) from each toy ensemble with the prediction in Eq.~(\ref{eq:c-prediction}); the high correlation (\(r=0.955\)) shows that the response-information ratio captures the dominant dependence of \(C\). Fig. \ref{fig:toy-mechanism}c illustrates the contrast with two representative ensembles selected from Fig. \ref{fig:toy-mechanism}a. In the scalar-dominated example, \(E\) remains strongly organized by \(\Psi\) (\(\operatorname{corr}(\Psi,E)=0.923\)) and the finite-response complementarity is weak (\(C=0.093\)). In the finite-response-complementary example, the scalar relation weakens (\(\operatorname{corr}(\Psi,E)=0.575\)) and the finite-response residual carries the missing ECS information (\(C=0.674\), \(\Gamma=1.997\)). The model therefore identifies a mechanism class for the CMIP6 behavior: forced-response heterogeneity that is poorly projected onto \(\Psi\), but appears in \(R\) with sufficient strength relative to transient kinetic contamination. Candidate climate realizations include cloud-feedback, pattern-effect, and ocean-heat-uptake differences. Additional model details and coordinate-level signatures are given in Supplementary Note 7 and Supplementary Fig. S6.

\subsection{Observation-facing HadCRUT5 diagnostic}

Finally, we apply the CMIP6-calibrated corrected coordinate in Eq.~(\ref{eq:corrected-estimator}) to the observable HadCRUT5 pair \((\Psi_\oplus,R_\oplus)\).

\begin{figure*}[t]
\centering
\includegraphics[width=\textwidth]{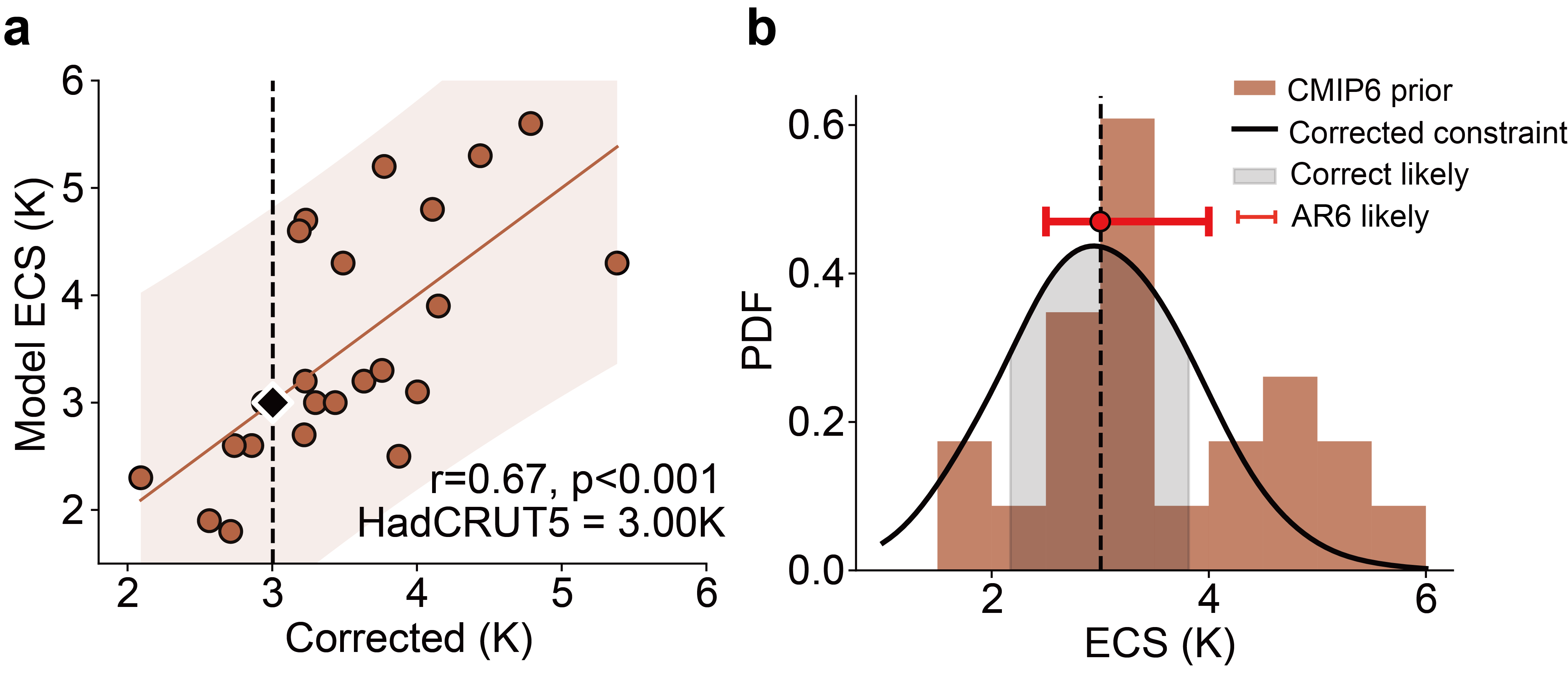}
\caption{\label{fig:hadcrut-correction} \textbf{Observation-facing finite-response diagnostic.} \textbf{a}, Corrected CMIP6 coordinate \(\hat E_{\rm corr}\) against model ECS, with HadCRUT5 marked. Colored shaded regions are model-calibration predictive bands. \textbf{b}, HadCRUT5 conditional density (black) and corrected 66\% diagnostic interval (gray), equal-weight CMIP6 ECS distribution (dark orange), and IPCC AR6 likely range and best estimate (red).}
\end{figure*}

Fig. \ref{fig:hadcrut-correction}a first checks the corrected coordinate inside CMIP6: \(\hat E_{\rm corr}\) is correlated with model ECS (\(r=0.675\), \(p<10^{-3}\)). Substituting the HadCRUT5 coordinates \((\Psi_\oplus,R_\oplus)\) into Eq.~(\ref{eq:corrected-estimator}) gives \(\hat E_{{\rm corr},\oplus}=3.004\,\mathrm K\), from \(E_0^{(6)}(\Psi_\oplus)=3.184\,\mathrm K\), \(R_0^{(6)}(\Psi_\oplus)=2.766\,\mathrm K\), \(R_\oplus=2.566\,\mathrm K\), and \(\Gamma_6=0.899\). Fig. \ref{fig:hadcrut-correction}b converts this corrected coordinate into a HadCRUT5 conditional diagnostic density: Monte Carlo samples of \(\hat E_{{\rm corr},\oplus}\) are passed through the CMIP6 conditional density \(p_6(E|\hat E_{\rm corr})\), marginalizing over HadCRUT5 rolling-window uncertainty in \(\Psi_\oplus\), ARX response-fit uncertainty in \(R_\oplus\), and CMIP6 calibration-bootstrap uncertainty in \(E_0\), \(R_0\), and \(\Gamma\) (Data and Methods; Supplementary Note 8). The resulting 66\% conditional diagnostic interval is 2.17--3.83 K, overlapping the IPCC AR6 likely range shown for reference. Supplementary Fig. S7a repeats the calculation as the common fluctuation window and response horizon are varied from 45 to 75 years, and Supplementary Figs. S7b,c collect additional HadCRUT5 sensitivity checks.

\section{Discussion and Conclusions}

This study shows that the CMIP6 weakening of a scalar historical-variability constraint can reveal, rather than simply remove, response information. The FDT-motivated intuition that fluctuations encode response must be interpreted here as a projected-closure statement: a finite historical GMST coordinate is not a full-state equilibrium correlation function. When that projection is incomplete, susceptibility information may reappear in a complementary finite-response coordinate. CMIP5 is fluctuation-memory dominated, whereas CMIP6 is finite-response complementary: after conditioning on \(\Psi\), CMIP6 retains a finite-response residual that organizes the ECS residual. The finite CO$_2$ response therefore does not replace \(\Psi\); it measures the part of the forced response not resolved by the scalar fluctuation-memory projection.

Operationally, this interpretation leads to a residual diagnostic. After the scalar dependence on \(\Psi\) has been removed, a finite-response coordinate is informative only if \(\Delta R\) organizes the remaining ECS residual \(\Delta E\). Annual GMST and common forcing histories do not uniquely identify the hidden coordinate, so the same residual-complementarity diagnostic should be extended to richer observables, including spatial patterns, ocean heat uptake, cloud-regime diagnostics, and single-forcing ensembles. More generally, emergent constraints on susceptibilities in nonequilibrium projected systems should combine marginal relationships with tests among complementary projected coordinates.

The result also suggests a model-ensemble design principle. The value of ensemble spread lies not in its size alone, but in whether it spans physically interpretable response directions that could contain the real system. A scalar fluctuation-response coordinate may be adequate when the relevant processes are represented and the forced-response differences not projected onto \(\Psi\) are small, or are themselves organized by \(\Psi\). If an ensemble spans a broad forced-response direction that is not represented by \(\Psi\) while still bracketing the real system, the weakened scalar constraint provides the contrast needed to identify complementary response coordinates and estimate the residual operator. Useful spread is therefore targeted and process-oriented, for example through controlled variations in cloud feedbacks, pattern effects, and ocean heat uptake, rather than indiscriminately large.

See the Supplemental Material \cite{SM} for methodological details, robustness checks, and additional figures and tables.

\begin{acknowledgments}
This work was supported by the National Natural Science Foundation of China (grant no. 12305044) and the National Key Research and Development Program of China (grant no. 2023YFE0109000).
\end{acknowledgments}

\section*{Author Contributions}
Y.Z. conceived the study and supervised the project. Z.Z. and Y.Z. developed the analysis, interpreted the results and wrote the manuscript. Both authors reviewed and approved the final manuscript.

\section*{Conflict of Interest}
The authors declare no competing interests.

\section*{Data Availability}
The data that support the findings of this article are openly available \cite{bib23,bib24,bib25}. CMIP5 and CMIP6 model simulations were obtained from the Earth System Grid Federation archive (\url{https://metagrid.esgf-west.org/search}). The HadCRUT5 data are available on the official website (\url{https://www.metoffice.gov.uk/hadobs/hadcrut5/}). The Python code used to process the data and generate the results and figures is openly available through Zenodo \cite{bib26}.

\bibliography{references}

\end{document}